\documentclass[11pt]{article}
 \usepackage[a4paper,hmarginratio=1:1,vmarginratio=2:3,totalwidth=15.7cm,
  totalheight=21.6cm]{geometry}

\usepackage{amsmath}
\usepackage{amssymb}
\usepackage{amsfonts}
\usepackage{epsfig}

\newcommand{\cL}{{\cal L}}

\newcommand{\ra}{\rightarrow}
\newcommand{\be}{\begin{equation}}
\newcommand{\ee}{\end{equation}}
\newcommand{\bea}{\begin{eqnarray}}
\newcommand{\eea}{\end{eqnarray}}

\DeclareMathSymbol{\mg}{\mathrel}{symbols}{"1D}

\newcounter{oldcounter}
\addtocounter{equation}{1}
\begin{document}
 \begin{flushright}
{OUTP-0709P}
\end{flushright}

\thispagestyle{empty}
\vspace{4cm}
\begin{center}
{\Large {\bf Higher dimensional operators and their effects in\\
\bigskip
(non)supersymmetric models\footnote{
This  is based on  work done in collaboration with I. Antoniadis and
E. Dudas.}}}
\vspace{1.cm}

{\bf  D.~M. Ghilencea$^{\,}$}\\
\vspace{0.5cm}
 {\it $ $Rudolf Peierls Centre for Theoretical Physics, University of Oxford,\\
 1 Keble Road, Oxford OX1 3NP, United Kingdom.}\\
\end{center}
\begin{abstract}
\noindent
It is shown that a 4D N=1 softly broken supersymmetric theory with 
higher derivative operators in the Kahler or the superpotential part
of the Lagrangian and with an otherwise arbitrary superpotential, can 
be re-formulated as a  theory without higher derivatives but with additional 
(ghost) superfields and modified interactions. The importance of the 
analytical continuation Minkowski-Euclidean space-time for the UV
behaviour of such theories is  discussed in detail.
In particular it is shown that  power  counting for divergences in
Minkowski space-time does not always work in models with higher 
derivative operators.
\end{abstract}

\vspace{2.8cm}
\noindent
\hspace{0.5cm}
{\sl Based on talk presented at}

\noindent
\hspace{0.5cm}{\sl ``Supersymmetry 2007'' Conference,
 26 July - 1 August 2007, Karlsruhe, Germany.}

\newpage
\setcounter{page}{1}

\section{Introduction}

Higher dimensional operators play an important role in 
the study of physics beyond the Standard Model (SM) and
its minimal supersymmetric  version (MSSM).  These 
operators are a generic presence in  phenomenological 
studies and are also predicted in effective field theories of
compactification, which attempt to recover in the low energy limit
the SM or MSSM. 

Higher dimensional operators, derivative or otherwise, can 
dramatically affect physics near the scale where they become relevant
and also the UV regime of the theory.  While  non-derivative higher
dimensional operators have been studied in the past, 
the case of higher derivative operators is less investigated.
Such operators were studied in the context of Randall-Sundrum models
\cite{RS},  cosmology \cite{Armendariz}, phase transitions \cite{BMP},
supergravity \cite{Stelle}, string theory \cite{Eliezer},
 have  applications to regularisation in 4D \cite{Slavnov,Leon},
and effects in the UV \cite{Antoniadis}, are generated by loop
corrections in extra dimensions in 5D and 6D orbifolds \cite{nibbelink}.
In the following  we would like to show how one can address
the effects of higher derivative operators, while still including 
arbitrary higher dimensional (non-derivative) operators. 
To this purpose we show that one can re-write a theory with higher
dimensional (derivative or otherwise) operators as a theory with 
only two-derivatives with additional superfields and
higher dimensional non-derivative interactions. This will be done in 
the context of  general 4D N=1  supersymmetric models and the case
of softly broken supersymmetry will also be considered.

Before proceeding to the supersymmetric case which we discuss in
 Section 3, we  review  the case of non-supersymmetric  higher derivative
 operators. This case has been investigated in \cite{Hawking}
where a path integral quantisation of a theory with such operators
was presented in an Euclidean formulation of the theory.
Interacting theories with higher derivative operators involve the
presence of ghost states,  and this can bring in  unitarity
violation. However this presence is not something specific to such
 theories, and is actually quite common: for example
4D gauge theories regularised \`a la Pauli-Villars also have ghosts
 \cite{Slavnov,Leon}.
Also in string theory the world-sheet theory of D-branes has 
non-renormalisability problems similar to gravity/supergravity, 
and allows the presence  of higher derivatives (and thus ghosts). 
Returning to field theories  with higher derivatives, 
 it was shown   in \cite{Hawking} that in such theories the
vacuum to vacuum amplitude and therefore  Green functions are
well-defined and there is no exponential growth of the amplitude, 
provided that
the ghost fields are not asymptotic states (they can
 still  be present as loop states). The result is that one violates 
unitarity  (at the scale of higher derivative operators) but one 
cannot create ghosts in the final state.

There is in general some ambiguity in theories with higher dimensional
operators, associated with the analytical continuation
form Minkowski to Euclidean space-time. 
The presence of higher derivative  operators brings in 
additional poles in the complex plane, associated with ghost 
states, and this requires one  define a contour of integration
around these poles  compatible with that of second order theories.
  To show the importance of 
the analytical continuation we provide examples where it
  dramatically alters the UV behaviour of the theory. As a result, 
different contours of integration give a different
UV behaviour for the theory. One consequence  is  that, contrary to
the naive expectation, the power counting of divergences
 in the Minkowski space-time does 
not always work in such theories (whether supersymmetric or not). 
This is then questioning,  in some circumstances,
the UV regularisation role usually  attributed in the literature to 
 higher derivative operators. For details on the results 
presented  below see \cite{Antoniadis,Antoniadis:2007xc}.

\section{Higher dimensional operators:  non-supersymmetric case}

Let us review the situation in  the non-supersymmetric case.
Consider the 4D Lagrangian
\smallskip
\begin{eqnarray}\label{action_4}
\cL= -\frac{1}{2}\, \phi \,(\xi\,\Box^2+\Box +m^2-i\,\epsilon)\,\phi 
- f(\phi),
\end{eqnarray}

\medskip
\noindent
with $\xi\equiv 1/M^2_*>0$  and $M_*$ is the (high) scale of 
``new physics'' where the higher derivative
term becomes  important; $f(\phi)$ is a general function 
accounting for terms in the potential other than the mass
term and can include higher dimensional (non-derivative)
 operators\footnote{The metric convention is $(+,-,-,-)$ and 
$\Box=\partial_\mu \partial^\mu$.}.
 In (\ref{action_4}) one can change the basis to  \cite{Hawking} (see
 also \cite{Antoniadis})
\medskip
\begin{eqnarray}\label{transition}
\varphi_{1,2}=-\frac{(\Box +m_{\pm}^2\pm i\,\epsilon^*) 
\,\phi \,\sqrt \xi}{
  (m_+^2-m_-^2+2 i \epsilon^*)^{1/2}},
\end{eqnarray}
where we introduced
\medskip
\begin{eqnarray}\label{mpm}
m_{\pm}^2
\equiv\frac{1}{2 \xi}\Big[1 \pm  (1-4\,\xi\,m^2)^{\frac{1}{2}}\Big],\qquad
\epsilon^*\equiv\frac{\epsilon}{(1-4 \xi m^2)^{1/2}}
\end{eqnarray}

\medskip
\noindent
Assume in the following that
 $\Lambda^2\gg  M_*^2\gg m^2$, 
therefore $\epsilon^*\approx  \epsilon\ll 1$ and also
$m_+^2\approx 1/\xi$, $m_-^2\approx m^2$;
$\Lambda$ is the UV cut-off of the theory.
With this,  eq.(\ref{action_4})  becomes
\medskip
\begin{eqnarray}\label{action_4g}
\cL=-\frac{1}{2} \varphi_1 (\Box +m_-^2 -i \epsilon^*) \,
\varphi_1+\frac{1}{2} \varphi_2
(\Box+m_+^2+i\,\epsilon^*)\,\varphi_2
-f\big(\varphi_2-\varphi_1\big)
\end{eqnarray}

\medskip
\noindent
The field $\varphi_2$ corresponds to a ghost field 
(has negative kinetic term). 
As seen from  (\ref{transition}), original $\phi\propto \varphi_1-
\varphi_2$
is a mixing
of a particle and a ghost-like degrees of freedom. 
The result is that the original Lagrangian with four  derivatives 
is ``unfolded'' into a two-derivative Lagrangian.
In Section~\ref{susy} we shall see how to perform a 
similar procedure  in  the supersymmetric  case
\cite{Antoniadis:2007xc}.

Above we assumed a particle-like $i\,\epsilon$  prescription in 
the  propagator of the 4-derivative theory of eq.(\ref{action_4}), 
and  obtained from this  the corresponding propagators'
prescriptions in the two-derivative formulation of
eq.(\ref{action_4g}). These prescriptions can also be seen from:
\medskip
\bea\label{a1}
\frac{1}{-\xi\,\Box^2-\Box-m^2+i\,\epsilon}=
\frac{1}{(1-4\,\xi\,m^2)^{1/2}}
\Big[\frac{1}{-\Box-m_-^2+i\,\epsilon^*}-
\frac{1}{-\Box-m_+^2-i\,\epsilon^*}\Big]
\eea

\medskip
\noindent
In the rhs we obtained the same propagators and prescriptions 
as in (\ref{action_4g}), for the particle ($m_-^2$) and ghost 
($m_+^2$) fields.  A natural assumption was made
above that 
the familiar pole  prescription for a particle field in the 
two-derivative theory remains true in the presence of the 
higher derivative term. This is shown in the lhs of (\ref{a1}) and in
(\ref{action_4})
by the sign in front of $i\,\epsilon$. This is 
certainly true at low $p^2$, but at higher  scales this is less clear
because there is no path integral formulation of (\ref{action_4})
 in the Minkowski space-time, to fully clarify which 
analytical continuation one should take for a higher derivative 
theory.  In any case, with our prescription (\ref{action_4}), one 
finds that the propagators in (\ref{action_4g}), (\ref{a1})
 emerged with opposite (signs for the) $i\epsilon^*$ prescriptions. 

In conclusion, once a particular prescription in the
fourth order theory is made, the second order theory has a well-defined
analytical continuation from Minkowski to Euclidean space-time.
In principle, one could ask  that the ghost propagator (second
term) in the rhs in (\ref{a1})
have a similar sign in front of $i\,\epsilon^*$
as the particle field. That would give in the lhs a {\it
momentum dependent} \,
 $\,\,\epsilon=\epsilon^*(1+2\xi\Box)$. Such analytical continuation 
can be considered and corresponds to  different  physics.
Note that in this case, for fixed $\epsilon$,  the relation 
$\epsilon^*\ll 1$ does not remain valid for $p^2\sim 1/\xi$
 and poles would move in the complex plane, far from the real axis.

To examine some of the consequences of (\ref{action_4})
we  calculate   the one-loop  correction $\delta m^2$ to 
the mass of $\phi$, for an interaction $f(\phi)=\lambda\,\phi^4/4!$
Using either formulation (\ref{action_4})  or (\ref{action_4g}), one has
\begin{eqnarray}\label{delta-mass-M}
 -i\, \delta m^2& =& 
-i \lambda\, \mu^{4-d}\int \frac{d^d p}{(2\pi)^d}
\frac{i}{-\xi p^4 +p^2-m^2+i\epsilon}\nonumber\\[9pt]
&=&\frac{-i \lambda\,\mu^{4-d}}{\sqrt{1-4 \xi m^2}}
 \int  \frac{d^d p}{(2\pi)^{d}}
\bigg[
\frac{i}{p^2-m_-^2+i \epsilon^*}-\frac{i}{p^2-m_+^2-i \epsilon^*   }
\bigg]\nonumber\\[9pt]
&=&
\frac{-i \lambda\,\mu^{4-d}}{\sqrt{1-4 \xi m^2}}\int_{\bf E}
\frac{d^d p}{(2\pi)^d} \bigg[\frac{1}{p^2+m_-^2}
+\frac{1}{p^2+m_+^2}\bigg],
\end{eqnarray}

\medskip
\noindent
where the last integral is in the Euclidean space, showed
by the index {\bf E};  $\mu$ is the standard 
mass scale introduced by the DR scheme, and $d=4-\omega$,
$\omega\ra 0$. The result for $\delta m^2$ is
 quadratically divergent; this is in contradiction
to what  power-counting suggests, when applied to the first line
in (\ref{delta-mass-M}), that the integral
should only be logarithmic divergent. The difference is explained
by  the fact that the two propagators in the second line above
are Wick-rotated in opposite directions (corresponding to $\pm
i\,\epsilon$),  bringing in a relative minus sign.
Moreover, if regarded in two-dimensions ($d\ra 2$)
the above equation would appear as finite by power counting, 
while the result shows it is log divergent. 
The conclusion is simple:  power counting in Minkowski 
space-time does not always work in effective field theories. 
This is an example which shows  that higher derivative operators 
do not always improve the UV behaviour of the theory, 
as commonly  stated in the literature.

To understand better the above results, it is useful to discuss
the Euclidean picture. The Euclidean version of (\ref{action_4}) is
\medskip
\begin{eqnarray}\label{action_E}
\cL_E=\frac{1}{2}\,\phi \,(\,\xi \,\Box_E^2-\Box_E+m^2\,)\,\phi 
+f(\phi)
\end{eqnarray}

\medskip
\noindent
At one-loop, the Feynman rules of $\cL_E$ with  
$f(\phi)=\lambda\,\phi^4/4!$ give
\medskip
\begin{eqnarray}\label{E_res}
 - \delta m^2 = 
- \lambda \,\mu^{4-d}\! \int_{\bf E} \frac{d^d p}{(2\pi)^d}\!
\frac{1}{\xi p^4 +p^2+m^2}
=\frac{- \lambda\,\mu^{4-d}}{\sqrt{1-4 \xi m^2}}
 \int_{\bf E} \!\frac{d^d p}{(2\pi)^{d}}
\bigg[\frac{1}{p^2+m_-^2}-\frac{1}{p^2+m_+^2}\bigg]
\end{eqnarray}

\medskip
\noindent
where all $p^2$ are evaluated in Euclidean space, as shown by 
subscript ${\bf E}$. A similar result is found by
working in  $\varphi_{1,2}$ basis. It is obvious that
this result 
has for $d=4$ 
no quadratic divergence but only a logarithmic one, in agreement
with power-counting in (\ref{E_res}).
Also, eq.(\ref{E_res}) is different from the last line in
 (\ref{delta-mass-M}) which has an opposite sign between the 
 contributions from $m_-$ and $m_+$; 
the origin of this different result in Euclidean 
case is due to the analytical continuation from the Minkowski to the 
Euclidean space-time.
 To conclude, we provided an explicit example where 
the same operator brings a different UV of a one-loop correction
in the Euclidean and Minkowski formulations of a theory\footnote{
One could impose  the prescription in (\ref{a1}) for the lhs
be such as the ghost and particle in the rhs  have identical
prescriptions and thus similar Wick rotations when going to the Euclidean
space-time. 
The result
in  (\ref{delta-mass-M}) would then be as in Euclidean case
(\ref{E_res}). It would be interesting to know if such a (momentum dependent) 
prescription is preferred by a Minkowskian path integral formulation 
of the higher derivative theory.}. This discussion shows the importance
of the analytical continuation in effective field theories with higher
derivative operators.

From the above discussion one could  conclude that 
the UV behaviour of $\delta m^2$ in the Euclidean formulation 
is improved  compared to that in  Minkowski space-time,
and that it has  no quadratic divergences. As discussed in \cite{Antoniadis}
a counterexample  shows  that this is not true. Consider the addition
to the Lagrangian  in (\ref{action_E}) of 
 $\Delta \cL_E= -z\, \phi^2\,\Box\,\phi^2$ which  also has 
dimension six, just like $\xi \phi\,\Box^2\,\phi$, and
 should then be included.
Its one-loop  effect on the mass of $\phi$ is
\medskip
\bea
\delta m^2\Big\vert_{\Delta\cL_E}
\equiv \frac{-16 \,z\,\mu^{4-d}}{\sqrt{1-4\xi\,m^2}}
\int_{\bf E}\frac{d^dp}{(2\pi)^d}\,
\bigg[\frac{m_+^2}{p^2+m_+^2}-\frac{m_-^2}{p^2+m_-^2}\bigg]
\eea

\medskip
\noindent
This result is clearly  quadratic divergent.
The Minkowski counterpart
to this result is also quadratic divergent, regardless of 
the signs $\pm i\epsilon^*$ of the prescriptions  one adds to the above
denominators for analytical continuation.
 Therefore the Euclidean version of a  4D theory
with additional dimension-six operators 
has quadratic divergences.  The effect of 
this dimension-six operator
was not included in the analysis in \cite{Grinstein}.

From the previous discussion we can also conclude that in a 4D theory,
in the presence of higher derivative operators, the mapping of the
radiative corrections computed  in Euclidean and Minkowski formulations
is not always at the level operator-by-operator. Instead, as it was shown in 
 \cite{Antoniadis}, for a given
order in perturbation theory, {\it a set of} operators in the Minkowski 
formulation can provide the same correction to the mass as in the
Euclidean theory, for an appropriate choice of the couplings.

We conclude that a 4D theory with higher
dimensional (derivative) operators can have a two-derivative
description in Minkowski space-time, and the UV regime of such a theory
depends strongly on the analytical continuation to the Euclidean 
space-time. In particular we saw that, contrary to common statements,  
 theories with such operators do not always have an 
improved UV behaviour. As a result, the role of higher derivative 
operators as UV regulators in Minkowski space-time 
can be  questioned, in the absence of a specific 
analytical continuation.

\bigskip
\section{Higher dimensional operators: the supersymmetric case}
\label{susy}

In the following we extend the above discussion to  the 
case of (softly broken) supersymmetric models \cite{Antoniadis,
Antoniadis:2007xc}.
Consider for example the
case of a Wess-Zumino  model with an additional 
supersymmetric higher derivative operator:
\medskip
\begin{eqnarray}\label{WZaction}
\cL&=&\int d^4\theta \,
\Phi^\dagger \,\big(1+ \xi \,\Box\big)\,\Phi
+\bigg\{\int d^2\theta\, \bigg[\,
\frac{1}{2} \,m \Phi^2 + \frac{1}{3}\lambda
  \,\Phi^3\,\bigg]+ c.c.\bigg\}-m^2_{0}\,\phi\, 
\phi^*\nonumber\\[9pt]
&=&
F^*\, \big( 1+\xi {\Box} \,\big) \,F
- \phi^*\Box \,\big( 1+\xi {\Box} \,\big) \,\phi
+i \partial_\mu \bar \psi  \,\bar\sigma^\mu
\big( 1+\xi {\Box} \big)\,  \psi\nonumber\\[9pt]
&+&\bigg[ \,
\frac12 \, m \,\big( 2 \,\phi\,F -\psi\psi \big)+\lambda\, 
\big(\phi^2 \,F-\phi\,\psi^2\big)\,
+h.c.\bigg]-m^2_{0}\,\phi\, \phi^*
\end{eqnarray}

\medskip
\noindent
Notice that in the presence of the higher derivative term, 
$F$ and $\Box\phi$ are dynamical degrees of freedom.
Let us compute the one-loop correction to the mass of scalar field
$\phi$ after soft supersymmetry breaking by $m_0^2\,\phi\,\phi^*$.
For simplicity we present the result for a massless case ($m=0$).
The one-loop scalar (fermionic) contributions  $\Delta m^2_b$  ($\Delta m^2_f$)
 to the mass of $\phi$ are 
\medskip
\begin{eqnarray}\label{WZintegral2}
\Delta m^2_b&=& 4 \,\,(i\lambda)^2  \,\,\int \frac{d^d p}{(2\pi)^d}
\frac{-1}{(-\xi)\,\,
[\,p^2 (1-\xi\,p^2)- m_0^2+i\,\epsilon ]\,
 (p^2-1/\xi+i\,\tilde\epsilon \,\,)},\qquad 
%(\tilde\epsilon=-\epsilon)
\nonumber\\[10pt]
\Delta m^2_f&=&  2\,\, (-i \lambda)^2 \int \frac{d^d p}{(2\pi)^d}
\frac{2}{(-\xi)\,\,
[\,p^2 (1-\xi\,p^2)+i\,\epsilon ]\, (p^2-1/\xi-i\,\epsilon \,\,)}
\end{eqnarray}

\medskip
\noindent
The first propagator under the integral of $\Delta m_b^2$ is that 
of $\phi$ in the presence of the higher derivative, in agreement with
(\ref{action_4}), while the second is that of the auxiliary field $F$, 
and for which  a prescription $\tilde\epsilon=-\epsilon$
is fixed by supersymmetry. Adding the above
 contributions gives~\cite{Antoniadis}
\begin{eqnarray}\label{mmm}
\Delta m^2_{\varphi}&=&\!\!\frac{\lambda^2}{4\pi^2}\,\,
\bigg\{(3\,\xi \,m_0^2)\, 2 \Lambda^2  
-
m_0^2\,\ln(1+\Lambda^2/m_0^2)
+\cdots\bigg\}
\end{eqnarray}

\medskip
\noindent
where the dots account for corrections which are at most logarithmic
in $\Lambda$ where $\Lambda$ is the UV cutoff of the loop integrals. 
In the decoupling limit of 
higher derivative operators $1/\xi=M_*^2\gg \Lambda^2$
one obtains the usual logarithmic correction for $\Delta m^2_\varphi$,
while for a mass $M_*< \Lambda$ the one-loop 
correction is quadratic divergent;
this could be somewhat surprising since
power-counting in (\ref{WZintegral2}) would suggest the
result is finite! As discussed in the previous section, 
here is another  example  that power-counting  applied to 
 (\ref{WZintegral2}) gives a misleading result, 
since such method ignores the poles prescriptions of the
ghost degrees of freedom.
The conclusion is that, for the analytical continuation considered here,
the breaking of supersymmetry, although  done by terms which are
soft, does not remain soft in the presence of  supersymmetric 
higher derivative terms.
 This discussion shows the importance 
of analytical continuation Minkowski-Euclidean space-time in solving 
the hierarchy problem  in the presence of such (supersymmetric) operators.
It is possible to find a continuation such as no quadratic 
divergence is present; this happens if the factor  in $\Delta m^2_{b,f}$
containing $p^4$  is a product of two propagators with similar Wick
rotations (i.e. same sign for $i\epsilon$); 
this would then require a {\it momentum-dependent}
prescription in the higher derivative theory with all its consequences
(see non-susy discussion after eq.(\ref{a1})).  This discussion underlines 
the need for a  path integral formulation of the higher derivative theories in
Minkowski space-time\footnote{No such formulation is currently available.},
from which one  could obtain the prescriptions on rigorous mathematical grounds.

\medskip
The next step is to extend to the supersymmetric level, the earlier 
finding that  a non-supersymmetric theory with
higher derivative operators can be reformulated 
as a theory with two derivatives only and
modified interactions \cite{Antoniadis:2007xc}.
Consider for this a  Lagrangian which generalises that in 
(\ref{WZaction}) (no susy breaking):
\medskip
\begin{eqnarray}\label{LL5}
\cL&=&\int d^4 \theta\, \Big[
\Phi^\dagger \Big(1+\,\xi\,\Box\Big) \Phi +
S^\dagger S\,\Big]
+\bigg\{ \int d^2 \theta \,\,\, W(\Phi,S)
 +h.c.\bigg\}
\end{eqnarray}

\medskip
\noindent
where $S$ is some  additional chiral superfield and $W$ is a
superpotential whose exact form is not relevant in the following.
This Lagrangian can be re-written in a new fields basis with only
two-derivative operators present, similar to the non-supersymmetric case. 
To see how this works  first notice that
one can re-write  $\xi\Phi^\dagger\, \Box\,\Phi$ 
as  $-\xi\,(D^2 \Phi/4)\,(\overline
D^2\Phi^\dagger/4)$. The idea is to replace somehow $\overline D^2\Phi$
(and its  conjugate) by another  chiral (anti-chiral) superfield. 
This can be done by using a $2\times 2$  unitary
 transformation to a new basis $\Phi_{1,2}$, of the form:
$\Phi=a_1\, \Phi_1+a_2\, \Phi_2$ and
$ m^{-1}\, \overline D^2\,
\Phi^\dagger= b_1\, \Phi_1+b_2\, \Phi_2$
where $m$ is some mass scale in the theory (for example the mass of
the field or a vev) and  $a_1,a_2$ and $b_1,b_2$ form a  unitary matrix.
Since $\Phi$, $\overline D^2\Phi$ are not independent,
one must also introduce a  constraint to account for this. 
To this purpose, the 
Lagrangian is  modified by adding $\Delta \cL$ where
\medskip
\begin{eqnarray}\label{delta-L}
\Delta \cL&=&
\int d^2\theta\,
\,\Big[m^{-1}\,
(a_1^* {\overline D}^2\,\Phi_1^\dagger +a_2^* {\overline
    D}^2\,\Phi_2^\dagger)
- ( b_1 \,\Phi_1+b_2 \,\Phi_2 )\Big]\, \Phi_3\,m_* +h.c.,
\nonumber\\[10pt]
&=& 
-4 \int d^4\theta \,\Big[\frac{m_*}{m}\,
(a_1^*\Phi_1^\dagger +a_2^*\Phi_2^\dagger)
\,\Phi_3\,+h.c.\Big]-
\int\,d^2\theta\,( 
b_1 \,\Phi_1+b_2 \,\Phi_2 )\, \Phi_3\,m_* +h.c.\qquad
\end{eqnarray}

\medskip
\noindent
Here $m_*\equiv {\sqrt{\xi}\, m^2}/{4}$; its 
dependence on $\xi$ was determined by requiring the constraint 
be absent in the limit $\xi\!\ra\! 0$ and by using that each squared
(super)derivative  comes with a $1/4$ factor. $\Phi_3$ is a
new ``constraint'' superfield introduced by the holomorphic
$\Delta \cL$. The new, total Lagrangian is $\cL'=\cL+\Delta \cL$ from which
we recover the constraint via the eq of motion for $\Phi_3$.

The next step is to bring to a canonical form the kinetic terms in 
$\cL'$. 
This is done via a unitary transformation from 
$\Phi_{1,2,3}$ to new $\tilde\Phi_{1,2,3}$ and a subsequent 
rescaling of $\tilde\Phi_{1,2,3}$  to obtain
canonically normalised  Kahler terms. In all this the superfield $S$
is spectator. After some
calculations one finds for $\cL'$ in basis $\tilde\Phi_{1,2,3}$ and
for a small $\xi$ ($m^2\ll 1/\xi$)
%\medskip
\bea
\cL'&=&
\int d^4 \theta\,\,
\Big[
\tilde\Phi_1^{\dagger} \tilde\Phi_1
-\tilde\Phi_2^{\dagger} \tilde \Phi_2
-\tilde\Phi_3^{\dagger}\tilde\Phi_3
+S^\dagger S\,
\Big]
\nonumber \\[10pt] &+&
\bigg\{
\int d^2 \theta \,\bigg[
\,\frac{-1}{\sqrt\xi}
\,\tilde\Phi_2\,\tilde\Phi_3 +
W\big(\tilde\Phi_{2}-\tilde\Phi_1;\,S\big)\bigg]+h.c.
\bigg\},\label{genW}
\eea

\medskip
\noindent
The above result shows that one can re-write the initial 
Lagrangian with higher
derivative operators in the Kahler term, as a  Lagrangian 
with two derivatives only. The difference to the original
formulation (\ref{WZaction}), (\ref{LL5}) is the appearance of extra
(ghost) superfields $\tilde\Phi_{2,3}$, of negative kinetic terms.
Their presence is easily understood -  we initially had $F$ and 
$\Box\phi$ as dynamical fields, which, by supersymmetry, require the
presence of two extra superfields, in agreement with  
(\ref{genW}). The relation between the original
 $\Phi$ and  $\tilde\Phi_{1,2,3}$,
is found to be $\Phi=(\tilde\Phi_2-\tilde\Phi_1)$. 
According to this the original superfield $\Phi$ contains a piece 
of ``ghost''!

The result in (\ref{genW}) is the supersymmetric equivalent 
of (\ref{action_4g})
and ``unfolds'' a theory with higher dimensional (derivative)
supersymmetric operators into a two-derivative theory, 
in a manifest supersymmetric way.
The  technique applied to obtain (\ref{genW}) is
general and does not depend on the exact form of $W$ which can contain
 operators of $D>4$.
The method can also be generalised to the case
when more higher derivative terms are present, provided the
appropriate holomorphic constraints ($\Delta \cL$) are introduced.  
Under eventual iteration,
we can map a theory with higher derivative terms in the Kahler potential,
into one without such terms, but eventually with higher
dimensional non-derivative operators.

The analysis can be extended to the case when higher
derivative terms are present in the superpotential, 
following the same steps \cite{Antoniadis:2007xc}.
 For example  one can consider
\medskip
\begin{eqnarray}\label{oW}
\cL&=&\int d^4 \theta\,\Big[\Phi^\dagger  \Phi + S^\dagger S\Big]+
\bigg\{
\int d^2 \theta \,\Big[s \,\sqrt{\xi}\,\, \Phi\,\,\Box \,\,\Phi +
W(\Phi,S)\Big] +h.c.\bigg\}
\eea

\medskip
\noindent
where $s=\pm 1$;  $W$ is  not specified, and can contain 
higher dimensional operators. One then follows the same 
procedure: first replace $\overline D^2\Phi$ by a chiral superfield,
then  re-write the higher derivative term as a D-term, and then
follow the same steps as in the case of higher derivative Kahler
terms. Since the auxiliary field of $\Phi$ is not
dynamical,  the form of the Lagrangian in the new basis 
 has only one additional (ghost) superfield. The equivalent 
Lagrangian of the second order theory is then obtained:
\medskip
\begin{eqnarray}\label{www}
\!\cL'\!&=&\! \int d^4 \theta\,\,
\Big[\tilde\Phi_1^{\dagger} \tilde\Phi_1-
\tilde\Phi_2^{\dagger} \tilde\Phi_2 +S^\dagger\,S\, \Big] 
+ \bigg\{ \int d^2 \theta \,\,\bigg[\,
\, \frac{\tilde\Phi_2^2}{4\,s\sqrt\xi}
+W\big(\tilde\Phi_2-\tilde\Phi_1;S\big)\, \bigg]+h.c.\bigg\}\qquad
\eea

\medskip
\noindent
This shows that one can indeed re-write a supersymmetric theory with 
higher derivative terms in the Kahler or superpotential part of the
Lagrangian, as a two-derivative theory, eventually with additional 
higher dimensional (non-derivative) operators.

The above results can also be extended to the case of supersymmetry
breaking. If one added soft breaking terms to the original Lagrangian
of (\ref{WZaction}), they are easily re-written in the new basis by using
the relation between the old and new fields
$\Phi=\tilde\Phi_2-\tilde\Phi_1$. This step raises the legitimate
question whether the technique we employed, by introducing a
holomorphic constraint (\ref{delta-L}), remains valid in the presence of 
soft breaking. We checked this is indeed the case for specific cases
of soft breaking terms. To do so we computed the spectrum of the model
in the old (\ref{WZaction}) and new basis (\ref{www})
of fields, after soft supersymmetry breaking, to
obtain the same result \cite{Antoniadis:2007xc}. 
This was checked for a standard
superpotential of the form given in eq.(\ref{WZaction}).

\section{Conclusions}

We have stressed the role that analytical continuation
Minkowski-Euclidean space-time is playing  in 4D models with 
additional higher dimensional (derivative) operators.
To show this we provided detailed examples where it can 
dramatically alter the UV behaviour of the theory.
These examples also showed that  power-counting 
is not always  a reliable tool in examining the UV of loop integrals in 
Minkowski space-time, in the presence of such operators.
This  questions the UV regularisation role
of higher derivative operators in Minkowski formulations of such
theories, in the absence of a rather specific analytical continuation.

In the supersymmetric case, it was discussed  how a
theory with higher derivative operators in the kinetic term or in
the superpotential and with an otherwise arbitrary superpotential 
is equivalent to a theory with two-derivatives only, 
with additional superfields and modified interactions,
 and with  higher dimensional (non-derivative)  operators.  
The new formulation of the theory generalises a similar
 non-supersymmetric equivalence and  can prove useful for 
applications.

\end{document}